\documentclass[%
reprint,
amsmath,amssymb,
aps,
floatfix,
prl
]{revtex4-2}

\usepackage{graphicx}
\usepackage{dcolumn}
\usepackage{bm}
\usepackage{color, soul}

\begin{document}
\newcommand{\state}[3]{\textsuperscript{#1}#2\textsubscript{#3}}
\newcommand{\muWcm}{$\mu$W/cm\textsuperscript{2}}
\preprint{APS/123-QED}

\title{Quantum States Imaging of Magnetic Field Contours based on Autler-Townes Effect in Yb Atoms}

\author{Tanaporn Na Narong}
\author{Hongquan Li}
\author{Joshua Tong}
\affiliation{%
 Department of Physics, Stanford University, Stanford, CA 94305\\
}%

\author{Mario Due\~{n}as}
\affiliation{%
California State University, East Bay, Hayward, CA 94542
}%

\author{Leo Hollberg}
\altaffiliation[Also at ]{Department of Geophysics, Stanford University, Stanford, CA 94305.}
\email{leoh@stanford.edu}

\date{\today}

\begin{abstract}
An inter-combination transition in Yb enables a novel approach for rapidly imaging magnetic field variations with excellent spatial and temporal resolution and accuracy. This quantum imaging magnetometer reveals ``dark stripes" that are contours of constant magnetic field visible by eye or capturable by standard cameras. These dark lines result from a combination of Autler-Townes splitting and the spatial Hanle effect in the \state{1}{S}{0}-\state{3}{P}{1} transition of Yb when driven by multiple strong coherent laser fields (carrier and AM/FM modulation sidebands of a single-mode 556 nm laser). We show good agreement between experimental data and our theoretical model for the closed, 4-level Zeeman shifted V-system and demonstrate scalar and vector magnetic field measurements at video frame rates over spatial dimensions of 5 cm with 0.1 mm resolution. Additionally, the \state{1}{S}{0}-\state{3}{P}{1} transition allows for $\sim\mu$s response time and a large dynamic range (from microtesla to many tesla). 
\end{abstract}

\maketitle
\begin{figure}[ht]
    \centering
    \includegraphics[width=0.48\textwidth]{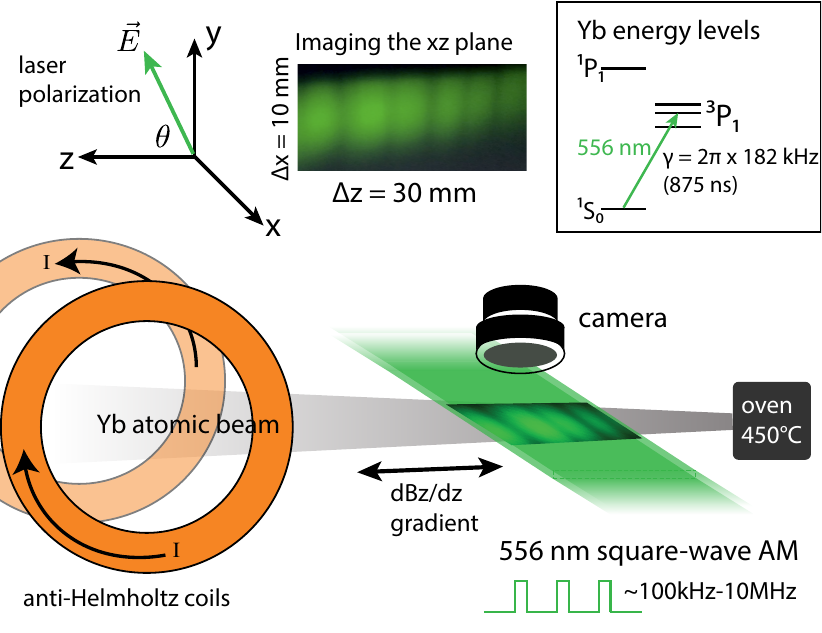}
    \caption{\label{fig:setup}Experimental system: a thermal atomic beam traveling horizontally (+$z$) is intersected by a 556 nm laser ($+x$) in resonance with the \state{1}{S}{0}-\state{3}{P}{1} transition in Yb and amplitude-modulated by an AOM. 
    Anti-Helmholtz coils produce \textbf{B} and $dB_z/dz$ primarily along the $z$ direction. 
    Imaging atomic fluorescence with the camera looking down the $y$-axis reveals a series of dark stripes in the $xz$ plane (10 mm $\times$ 30 mm), visible by eye and a simple camera. 
    The laser is linearly polarized at a rotatable angle $\theta$ to the $z$-axis. 
   } \end{figure}

A new type of atomic magnetometer that uses Yb quantum states was discovered in an experiment that employed stimulated optical forces to effectively manipulate atomic velocities and trajectories~\cite{narong2021stimulated}. This Yb magnetometer operates with a single 556 nm laser on the \state{1}{S}{0}-\state{3}{P}{1} transition (Fig.~\ref{fig:setup}). We observe dark stripes in bright green fluorescence, visible by eye in the presence of magnetic field gradients and multiple strong optical fields. These stripes, which represent contours of constant magnetic field, enable direct mapping of spatial gradients and nonuniform magnetic fields with rapid response time. The magnetometer can also measure the vector orientation of magnetic fields.

The dark stripes observed resemble those from Coherent Population Trapping (CPT)~\cite{arimondo_v_1996}, first observed in 1976 by Alzetta et al.~\cite{alzetta1976experimental}. Soon after, Tam \& Happer described imaging inhomogeneous magnetic fields in sodium vapor using optical-RF double resonance~\cite{tam1977optically}. However, the Yb dark stripes result from different physics----mainly Autler-Townes (AT) splitting~\cite{autler_stark_1955} combined with the spatial Hanle effect~\cite{hanle1924magnetische, novikov1975hanle}.
AT splitting in our system is produced by two or more ``strong" laser fields, where strong is relative to the low saturation intensity, $I_{sat} = 136$ \muWcm for the \state{1}{S}{0}-\state{3}{P}{1} transition driven in a V-configuration.
Although AT, CPT, and Electromagnetically Induced Transparency (EIT)~\cite{harris_electromagnetically1997, fleischhauer_electromagnetically_2005} are all quantum phenomena that can lead to decreased absorption and fluorescence in multilevel systems, refs.~\cite{abi-salloum_electromagnetically_2010, anisimov_objectively_2011, zhou2019analysis} show that CPT and EIT do not occur in closed V systems that have ground states with zero electronic and nuclear spin, such as even isotopes of Yb.  Other papers treat different configurations and come to different conclusions~\cite{vafafard_coherent_2016, rathod_magnetometry_2015, mompart2001coherent, singh2015observation}. Our observations and the model agree best with AT as the dominant mechanism (see SM~\cite{supplement}). 

To understand the physics of the dark stripes, we developed a numerical model of the 4-level, closed \state{1}{S}{0}-\state{3}{P}{1} transition driven by multiple strong optical fields. This model allows direct computation of magnetic field tomography from images collected in experiments as well as generation of fluorescence images given B-field input. The results and analysis are broadly applicable to other atoms with similar structures (e.g. Ca, Sr, Mg). The simple experimental setup in Fig.\ref{fig:setup} uses a 556 nm light sheet (3 mm $\times$ 30 mm) with multiple coherent sidebands intersecting a Yb thermal atomic beam. The fluorescence, captured by a camera, is sensitive to the magnetic field's projection relative to the laser polarization, which is linearly polarized at a rotatable angle $\theta$ to the $z$-axis. Large anti-Helmholtz coils along $z$ create a magnetic field gradient, primarily $dB_z/dz$. The inset image was taken with 2 MHz modulation and $y-$polarized light ($\theta = \pi/2$).

Yb is well-suited for imaging spatially and temporally varying magnetic fields due to compelling characteristics of the \state{1}{S}{0}-\state{3}{P}{1} transition.
With an intermediate transition lifetime (875 ns), the linewidth (182 kHz) is narrow enough to provide good magnetic field sensitivity ($\approx$1 nT for a 3 ms image). 
The low $I_{sat}$ allows mW laser powers to generate strong-field AT splitting and high dark-line contrast. 
The short lifetime also ensures good spatial resolution (around 0.1 mm longitudinally) and enables recording of fast ($\sim\mu$s) magnetic field dynamics. 
The 556 nm green wavelength is ideal for imaging (see SM~\cite{supplement} for example videos). 
These features distinguish this Yb magnetometer from others, offering new and complementary applications.  
This system achieves rapid, spatially-resolved B-field tomography over a wide range of fields, from $\mu$T to many Ts. 
We show later that the magnetic field at each dark stripe can be computed accurately and directly using known g-factors and laser sideband characteristics.

Beyond scalar measurements, our Yb magnetometer can measure the B-field vector orientation because both the AT splitting and the spatial Hanle effect are sensitive to the laser polarization \textbf{E}.
The observed fluorescence is highly dependent on the relative orientation of \textbf{E}, the local magnetic field vector \textbf{B}, and the observation direction. 
Some EIT~\cite{cox_measurements_2011, mckelvy2023application} and CPT~\cite{pradhan_polarization_2012} vector magnetometers also have this capability.
As shown later, in some cases, even a single laser beam with a fixed polarization can determine all three components of \textbf{B} from a fluorescence image using forward simulations and parameter space searches based on our model.

The market for atomic magnetometers (Cs, Rb, K, He, proton, NV-diamond) is much smaller than that for solid-state devices (Hall, GMR, fluxgates), but atomic magnetometers offer excellent sensitivity and accuracy. 
Many high-performance devices rely on optical pumping or CPT, detected via optical absorption or polarization rotation. Significant advances are summarized in these excellent reviews~\cite{budker_resonant_2002, budker_kimball_2013, grosz_haji-Sheikh_mukhopadhyay_2018, fagaly_superconducting_2006, jensen_magnetometry_2017, budker_optical_2007-1, mitchell2020colloquium, lu2023recent, brajon2023benchmark}. 
Most relevant for comparative context are magnetometers capable of real-time imaging of magnetic fields (scalar or vector) or rapid tomography~\cite{tam1977optically, castellucci_atomic_2021, vengalattore_high-resolution_2007, yang_scanning_2017, courteille_magnetic_2001, nielsen_characterization_2008, qiu_visualization_2021, lu2023twodimensional, jackson_magneto-optical_2019, lu2023two, hunter2023free, mikhailov2009magnetic, lu2023two, alem2017magnetic}. 
Some of these systems utilize BECs~\cite{vengalattore_high-resolution_2007, yang_scanning_2017, wildermuth2006sensing} or cold atoms~\cite{nielsen2008characterization, eliasson2019spatially} to achieve excellent spatial resolution and sensitivity.
Vapor cells enable diverse applications, including absorption imaging with vortex laser beams for 3D vector measurements~\cite{castellucci_atomic_2021}, gradiometers~\cite{sheng2017microfabricated, lucivero2022femtotesla, deans2021electromagnetic, lucivero2022femtotesla, campbell2022gradient}, and arrays~\cite{boto2017new, borna2017twenty, coussens2024modular}. 
More similar examples to ours include B-field gradient imaging in Cs vapor cells using optical-microwave double-resonance~\cite{fescenko2014imaging} and imaging of dark spots in Sr fluorescence on the singlet \state{1}{S}{0}-\state{1}{P}{1} transition near zero magnetic field~\cite{jackson_magneto-optical_2019} due to the spatial Hanle effect, but with weak fields and no AT effect. We adapted their calculation of directional fluorescence in our model.
Avan and Cohen-Tanoudji analyzed the Hanle effect in the 4-level V-system with strong fields~\cite{avan1975hanle}.

\begin{figure}[b]
\centering
\includegraphics[width=0.47\textwidth]{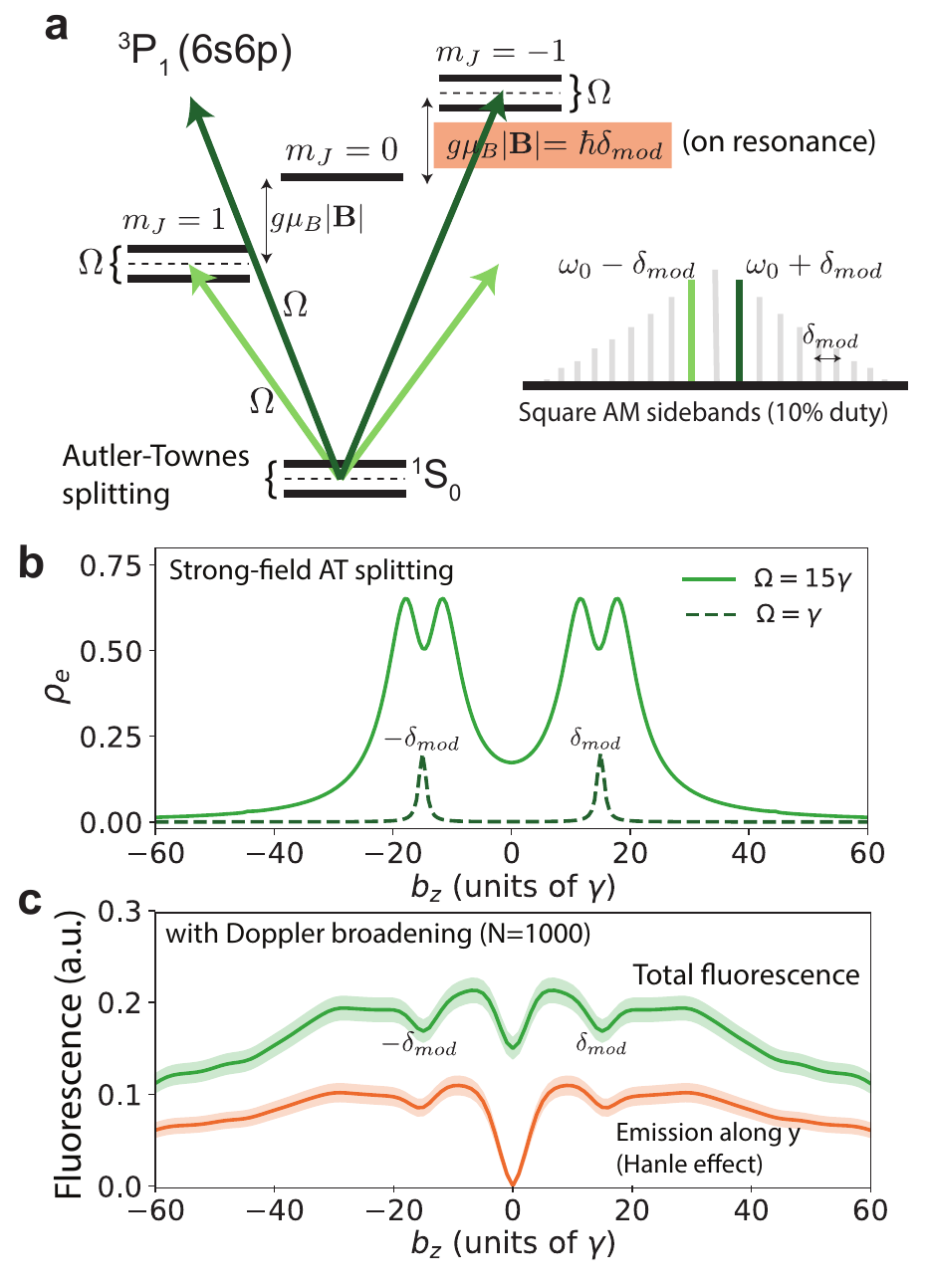}
\caption{\label{fig:levels_AT_fluorescence}(a) Energy levels of the Yb \state{1}{S}{0}-\state{3}{P}{1} transition showing AT splitting in three states when driven by strong resonant fields with equal Rabi frequency $\Omega$.
Our initial model includes only the first harmonic sidebands at $\omega_0 \pm \delta_{mod}$ and B-fields along $z$. $y$-polarized fields drive $\sigma^\pm$ transitions to $m_J=\pm1$ states, and resonances occur at $b_z = \pm \delta_{mod}$.
(b) AT splitting caused a decrease in fluorescence ($\propto$ total excited-state population $\rho_e$) on resonance with strong driving fields ($\Omega = 15\gamma$) but not with weak fields ($\Omega = \gamma$).
(c) With strong fields, Doppler-broadened fluorescence has three minima at $b_z = 0$ and $\pm \delta_{mod}$. With the spatial Hanle effect included (bottom curve), the fluorescence emitted along $y$ goes to zero where $b_z=0$. Uncertainty bands represent one standard deviation for Monte Carlo simulations with $N$ atoms.
} 
\end{figure}

\textit{Model--} We consider the closed, 4-level \state{1}{S}{0}-\state{3}{P}{1} transition (Fig.~\ref{fig:levels_AT_fluorescence}(a)) in even isotopes (most abundant \textsuperscript{174}Yb) with zero nuclear spin ($I=0$) and no hyperfine structure or complications of optical pumping. 
The \state{3}{P}{1} state consists of Zeeman states $m_J=-1,0,1$, with corresponding energy shifts $m_J g \mu_B |\textbf{B}|$ (in the low field regime), with  $g=1.49282(5)$~\cite{martin1978atomic, budick1967gj, baumann1968gj} and $\mu_B$ the Bohr magneton. In our setup, this transition is driven by amplitude-modulated light at modulation frequency $\delta_{mod}$, which consists of a carrier at $\omega_0$ (set to the resonance frequency of the $m_J=0$ state) plus a series of evenly spaced sidebands at $\omega_0 \pm n\delta_{mod}$, where $n = 1, 2, 3,...$

The model initially included only the two first-harmonic sidebands at $\omega_0 \pm \delta_{mod}$ of the laser polarized along $y$ ($\theta = \pi/2$) perpendicular to the imaging plane, with the quantization axis along $z$ and no magnetic field along $x$ and $y$ directions (Fig.~\ref{fig:levels_AT_fluorescence}(a)). 
The $y-$polarized light can be decomposed into a combination of LCP and RCP light that drive $\sigma^+$ and $\sigma^-$ transitions to the $m_J = \pm1$ states, forming a Zeeman-shifted V-system. 
With symmetric detunings and Zeeman shifts, resonances occur at $g\mu_B|\textbf{B}| = \hbar\delta_{mod}$. 
In this configuration, strong-field AC Stark effects cause the three coupled states to split (AT splitting)~\cite{autler_stark_1955}, which leads to decreased absorption and fluorescence on resonance.

For laser polarization $\theta$ and arbitrary \textbf{B}$=(B_x, B_y, B_z)$, the Hamiltonian of this 4-level atom is
\begin{equation}\label{H_theta_xyz}
    H_\theta= 
    \hbar \begin{pmatrix}
    0 & \Omega_{\sigma}^*(t)\sin{\theta} & \Omega_{\pi}^*(t)\cos{\theta} & \Omega_{\sigma}^*(t)\sin{\theta}\\
    \Omega_{\sigma}(t)\sin{\theta} & \omega_0 + b_z & b_{xy}^* & 0 \\
    \Omega_{\pi}(t)\cos{\theta} & b_{xy} & \omega_0 & b_{xy}^* \\
    \Omega_{\sigma}(t)\sin{\theta} & 0 & b_{xy}^* & \omega_0 - b_z
    \end{pmatrix}.
    \end{equation}
Here, $b_z = g\mu_B B_z/\hbar$ denotes the Zeeman (frequency) shift due to the $B_z$ field component. Similarly, $b_{xy}$ encapsulates the remaining magnetic field components $B_x$ and $B_y$ where $b_{xy} = \frac{g\mu_B}{\hbar\sqrt{2}}(B_x + iB_y)$. Strong electric dipole interactions produce the time-dependent matrix elements of $H_{\theta}$. $\Omega_{\sigma}$ and $\Omega_{\pi}$ denote contributions that produce $\sigma$ transitions to the $m_J=\pm1$ states and $\pi$ transitions to the $m_J=0$ state, respectively, and are defined as follows.
    \begin{align}
    \Omega_{\sigma}(t) &= \frac{i}{\sqrt{6}}(\Omega_1\cos{(\omega_1 t)} + \Omega_2\cos{(\omega_2 t)}) \\
    \Omega_{\pi}(t) &= \frac{1}{\sqrt{3}}(\Omega_1\cos{(\omega_1 t)} + \Omega_2\cos{(\omega_2 t)}),
    \end{align}
where $\Omega_{1,2}$ are the Rabi frequencies and  $\omega_{1,2} = \omega_0 \pm \delta_{mod}$ are the frequencies of the first-harmonic sidebands. The pre-factors $i/\sqrt{6}$ and $1/\sqrt{3}$ are Clebsch-Gordan coefficients. 

With the Hamiltonian, we calculate the total time-averaged fluorescence by summing the excited-state population from the density matrix $\rho$, which is obtained by numerically solving the Lindblad Master Equation using the Python package QuTiP~\cite{johanssonQuTiPOpensourcePython2012, johanssonQuTiPPythonFramework2013}.
The code is available on GitHub~\cite{nanarong2024githubYbmag} with a documentation provided in TN's thesis~\cite{narong2023thesis}.
In the calculation we make the rotating-wave approximation as the modulation frequencies (100 kHz-10 MHz) and weak-field Zeeman shifts ($\approx$2.1 MHz/G = 21 kHz/$\mu$T) are small relative to $\omega_0$.

In the simplest case where the magnetic field is along $z$ ($b_{xy}=0$) and $\theta=\pi/2$, only $\sigma$ transitions are allowed. Assuming equal Rabi frequencies ($\Omega_1 = \Omega_2 = \Omega$) and with $\delta_{mod}=15\gamma$, Fig.~\ref{fig:levels_AT_fluorescence}(b) shows the total excited-state population $\rho_e$ as a function of Zeeman shift $b_z$ for weak and strong fields. 
On the resonances at $b_z = \pm \delta_{mod}$ where the Zeeman shifts match the field detunings, only the strong fields ($\Omega = 15\gamma$) produced apparent Autler-Townes splitting and a decrease in fluorescence on the resonances. 
These fluorescence ``dips" remain visible in the presence of Doppler broadening due to atomic velocities $v_x$ in the direction of laser propagation (Fig.~\ref{fig:levels_AT_fluorescence}(c)). 
We included this effect by averaging $\rho_e$ across $N=1000$ transverse velocities $v_x$ sampled from a Gaussian distribution with zero mean and standard deviation of 4.5 m/s (FWHM $\approx$10.6 m/s) estimated from the atomic beam divergence. The Doppler broadened curve has three minima at $b_z = 0, \pm \delta_{mod}$, which would result in three dark stripes on a fluorescence image. At $\Omega = 15\gamma$, the FWHM of the dips is around $ 5\gamma \approx 900$ kHz, which is significantly narrower than the Doppler width (19 MHz). The uncertainty band shows the standard deviation $\sqrt{\bar\rho_e(1-\bar\rho_e)/N}$ for a Monte Carlo simulation with $N$ atoms.
    
While the computed total fluorescence accurately predicts the location of the dark lines, $\rho_e$ does not account for the system geometry, where the camera only collects fluorescence emitted within in a cone around the $+y$ direction. To improve predictions of lineshapes and contrast, we incorporated the spatial Hanle effect to account for the spatial distribution of the dipole radiation.
This effect was carefully studied by Jackson \& Durfee for the broad \state{1}{S}{0}-\state{1}{P}{1} transition in Sr driven by weak fields~\cite{jackson_magneto-optical_2019}. 
We adapted their analysis to compute the fluorescence emitted in the $y$-direction ($I_y$) from the density matrix (see SM~\cite{supplement} for the derivation), resulting in the bottom curve in Fig.~\ref{fig:levels_AT_fluorescence}(c). The Hanle effect did not alter the locations of the dark lines but the signal magnitude roughly halved, and $I_y$ went to zero at $b_z = 0$, as expected for $y$-polarized electric dipole moments. This correction improved agreement with our experimental results, but some discrepancies remained.

Next, we generalize the dark-resonance conditions for more than two driving fields and for 3D magnetic fields. 
First, we note that, with Doppler broadening, any two sidebands in the square wave harmonics (Fig.~\ref{fig:levels_AT_fluorescence}(a)) produce dark resonances when their relative detuning (an integer multiple of $\delta_{mod}$ matches the Zeeman shifts between the $m_J = \pm1$ states ($2g\mu_B|\textbf{B}|$). Hence, the dark-resonance condition becomes
\begin{equation}\label{eq:dark-line-condition}
\frac{g\mu_B}{\hbar}|\textbf{B}| = \frac{n}{2}\delta_{mod},
\end{equation}
where $n= 1, 2, 3, ...$. ($n=2$ recovers the base case because the two first-harmonic sidebands are $2\delta_{mod}$ apart). Equation~\eqref{eq:dark-line-condition} now describes a series of contours of constant magnetic fields where the fluorescence goes ``dark" and enables direct computation of $|\textbf{B}|$ at the location of each dark stripe based on the known $\delta_{mod}$ and the g-factor. 
To verify this relationship and better reproduce the experimental results, we extended the model to include 7 strong fields in total: the carrier and evenly spaced sidebands up to the third harmonics. We also analyzed the dark resonances explicitly in the 3D picture, by expressing $|\textbf{B}|$ in terms of its vector components and investigating the effects of $B_x$ and $B_y$ on the fluorescence pattern (see End Matter). Forward modeling of $I_y$ enabled us to estimate all components $|B_x|$, $|B_y|$, and $|B_z|$ at small fields (below 200 $\mu$T) from experimental data as shown below.
 
\begin{figure}[b]
\centering
\includegraphics[width=0.45\textwidth]{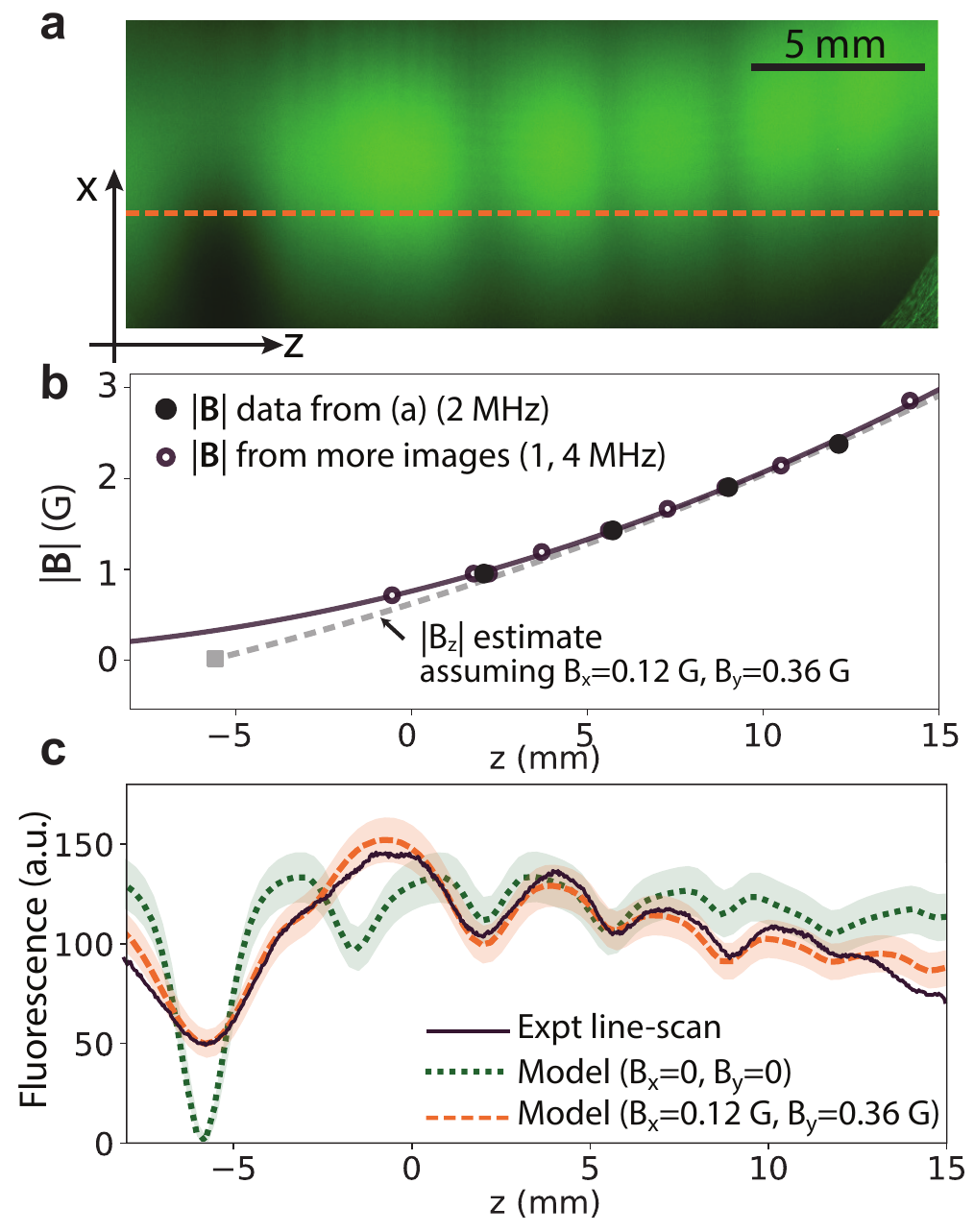}
\caption{\label{fig:B-field-mapping}B-field tomography from (a) the fluorescence image with dark stripes that correspond to contours of constant magnetic fields. The darkest spot near $z = -5$ mm is where $B_z = 0$. The laser beam was $y$-polarized and modulated at 2 MHz. The dashed line marks where the line-scan in (c) was taken.
(b) $|\textbf{B}|$ calculated at the dark-stripe locations extracted from (a) and additional frames taken at 1 and 4 MHz (error bars too small to show). $|B_z|$ (dashed line) was calculated based on the estimates $B_x=0.12$ G, $B_y=0.36$ G, determined by fitting the model to the line-scan in (c). Both curves are quadratic least square fits.
(c) Intensity line-scan taken from (a) and theoretically calculated fluorescence assuming $B_x = B_y = 0$ (green dotted line) and with $B_x = 0.12$ G, $B_y = 0.36$ G (orange dashed line), assumed uniform across $z$. Each uncertainty band represents the standard deviation.}
\end{figure}

\textit{Experiment--} Starting with a fluorescence image in Fig.~\ref{fig:B-field-mapping}(a) with a series of vertical dark stripes resulting primarily from a $B_z$ field with a $dB_z/dz$ gradient generated by the large coils, we extracted the dark-line locations and computed the field magnitude $|\textbf{B}|$ with Eq.~\eqref{eq:dark-line-condition}, plotted in Fig.~\ref{fig:B-field-mapping}(b) with a quadratic least-square fit. According to the Hanle effect, the darkest spot near $z$=$-5$ mm indicates where $B_z = 0$. The laser was square-wave AM modulated at 2 MHz at 10\% duty cycle and $y$-polarized. Additional data points were extracted from images taken with 1 MHz and 4 MHz modulation. From these three frames (3 ms), we estimate the uncertainty in $|\textbf{B}|$ to be $\approx$ 1 $\mu$T (0.01 G) based on the width of the dark lines and signal-to-noise ratio. This precision can be improved with more frames. The thin ``sheet" of light in the $xz$-plane does not capture variations in \textbf{B} along the $y$-axis (none expected on that scale). This assumption could break down if variations of \textbf{B} in the $y$-direction over the sheet's thickness $\Delta y$ is greater than the power-broadened transition linewidth, in which case it would blur out the dark lines.
In our setup, $\Delta y=3$ mm and the power-broadened (dark) linewidths range from 360 kHz to 2 MHz, corresponding to $\Delta B\approx$ 7-100 $\mu$T (70 mG to 1 G) taking into account both $\Delta m_J = 1,2$ V-transitions. 

Next, we investigate the agreement between our theoretical fluorescence and experiment. Using the computed $|\textbf{B}|$, we first generated a theoretical fluorescence curve by assuming $B_x$=$B_y$=0 and that the B-field gradient was solely $dB_z/dz$ (green curve). This baseline calculation produced the green dotted curve in Fig.~\ref{fig:B-field-mapping}(c) with the uncertainty band (one standard deviation). We see two significant discrepancies from the experiment (black curve). First, the minimum fluorescence near $z=-5$ mm (where $B_z=0$) is well above zero even after background subtraction. This indicates that $B_x\neq0$ because $B_x$ causes dipole precession about the $x$-axis (away from the polarization axis $y$), allowing some fluorescence to reach the camera. Second, the apparent shifts and distortion near the first dip suggest that $B_y \neq 0$. 
With $y-$polarized light, $B_y$ effectively shifts the locations of the dark lines (contours of constant $|B|$) without lifting the fluorescence minimum from zero because $B_y$ does not change the dipole orientation, an important distinction from $B_x$ (see End Matter for more details and illustrations in Figs.~\ref{fig:contour_bxbz} and~\ref{fig:contour_bybz}.)
Using our model and interactively calculating the fluorescence (see End Matter) for a range of constant $B_x, B_y$ values, we found that $|B_x|=0.12(2)$ G, $|B_y=|0.36(2)$ G produced the theoretical curve (orange dashed line) that best agrees with the experiment (black curve). These values are consistent with independent Hall probe measurements of the Earth's field outside the vacuum chamber, and within uncertainties of the Hall sensor. With these estimates, we computed revised values of $|B_z|$, as shown in Fig.~\ref{fig:B-field-mapping}(b).

\textit{Discussion--} It was initially surprising that a single sheet of $y$-polarized light in the $xz$-plane provides information about both the field magnitude and direction, including variations of $B_x$ and $B_z$ in the $xz$-plane (assuming no significant variation of \textbf{B} in the thin (3 mm) $y$-direction).
The additional vector information comes from the spatial Hanle effect at small fields, where fluorescence redistributes due to the magnetic moment's precession in local fields increasing away from \textbf{B}=0~\cite{jackson_magneto-optical_2019, narong2023thesis}. (At large fields, the precession produces high-contrast Zeeman quantum beats in fluorescence, observed for this Yb transition~\cite{lowe1989zeeman}.)
Similar to EIT~\cite{cox_measurements_2011} and CPT~\cite{pradhan_polarization_2012} magnetometers, we gain more vector information by rotating the laser polarization, as discussed in~\cite{narong2023thesis}, and 3D measurements of \textbf{B} are straightforward by scanning the light sheet in the $y$ direction.
A detailed exploration of these vector capabilities is left for future work.

Most atomic magnetometers (Rb, Cs, K, He) use small vapor cells (mm-cm), with high atomic densities (10\textsuperscript{10}-10\textsuperscript{12}/cm\textsuperscript{3}) that measure uniform fields via optical absorption or polarization rotation~\cite{budker_kimball_2013, grosz_haji-Sheikh_mukhopadhyay_2018, fagaly_superconducting_2006, jensen_magnetometry_2017, budker_optical_2007-1, mitchell2020colloquium, lu2023recent}.
In contrast, this Yb magnetometer relies on Autler-Townes splitting for all magnetic field magnitudes and the spatial-Hanle effect at low fields. 
This combination enables exciting imaging and vector capabilities as demonstrated, with measurement bandwidths from DC to $\approx$500 kHz and no dead-time.
It can measure fields from $\mu T$ to $>$10 T (see SM~\cite{supplement}), though we have not tested beyond 20 mT (200 G).
The atomic density (up to 10\textsuperscript{10}/cm\textsuperscript{3}) is low enough to avoid collision limitations, and optical pumping does not occur in even isotopes.

Our current setup effectively demonstrates the concepts and capabilities of the Yb quantum imaging magnetometer, but it is not designed, nor optimal, for magnetometry. Yb has several unique capabilities, but also limitations. It requires high temperatures ($\gtrapprox$400 $^\circ$C) due to its low vapor pressure, making vapor cell production challenging.
The ultimate sensitivity is lower than typical atomic magnetometers due to the broader linewidth (182 kHz vs $\approx$0.1-1 kHz) and lower atomic density. 
Additionally, at higher fields, signals from other Yb isotopes may need to be addressed.
However, our current system produces bright green fluorescence images in 3 ms (camera frame rate limited), with $>$ 10,000 spatially resolved B-field measurements over the 10$\times$30 mm\textsuperscript{2} interaction region.
Using five cameras, a spatial range of 1 m with 0.1 mm resolution would be feasible (see SM~\cite{supplement}).
The estimated sensitivity is $\approx$1 nT for a 3 ms image, with a field uncertainty of $\approx$100 nT~\cite{narong2023thesis}.

Our approach does not require a well-collimated atomic beam and works best with Doppler averaging. Fluorescence is detected from all atoms within the homogeneous excitation width, including modulation sidebands.
Power-broadened linewidths of the dark stripes ranged from 360 kHz to 2 MHz (FWHM), much narrower than the $\approx$19 MHz transverse Doppler width. This Yb system operates effectively in an unshielded environment and opens up possibilities for new applications complementary to existing magnetic sensors, covering a wider range of spatial, temporal, and field parameters.

\begin{acknowledgements}
We are grateful to many who have contributed to our understanding of this Yb magnetometer, including Stuart Ingleby, Erling Riis, and the B-sensing group at Strathclyde University, Svenja Knappe, John Kitching, Benjamin Lev, Sonja Franke-Arnold, Jason Hogan, and Megan Nantel. Our initial discovery of this effect was with ONR support, grant N000141712255; followed by the NSF QLCI OMA-2016244. 
Some of the computing was performed on the Sherlock cluster. We thank Stanford University and the Stanford Research Computing Center for providing computational resources and support. HL was supported by a Stanford Bio-X SIGF fellowship.
MD was supported by the NSF QLCI and Cal-Bridge. JT was supported by an EDGE fellowship and the Stanford physics department fellowship. We appreciate valuable assistance with the experiments by summer students: Luz Elena Martinez, Daniel Kuelbs, Judy Liu, and Zhengruilong Wang.
\end{acknowledgements}

TN developed the theory and numerical model in Python and led the data analysis efforts. HL initially discovered the dark stripes, developed an initial model and collected experimental data with help from JT and MD. TN prepared the manuscript with significant input from LH. LH guided all aspects of this project.

\bibliography{mybib, leoh-magnetometer}
\nocite{bai1986transient}

\appendix
\section{End Matter}

\textit{Appendix: $B_x$ and $B_y$ field components alter the observed fluorescence differently--} Our model enables direct computation of fluorescence as a function of an arbitrary magnetic field $\textbf{B} = (B_x, B_y, B_z)$. As we generalize the dark resonance condition~\eqref{eq:dark-line-condition} from the 1D ($B_z$-only) to the full 3D picture, fluorescence contour plots as a function of two varying components of \textbf{B} are useful tools for visualizing the effects of different field components. 

\begin{figure}[ht]
\centering
\includegraphics[width=0.49\textwidth]{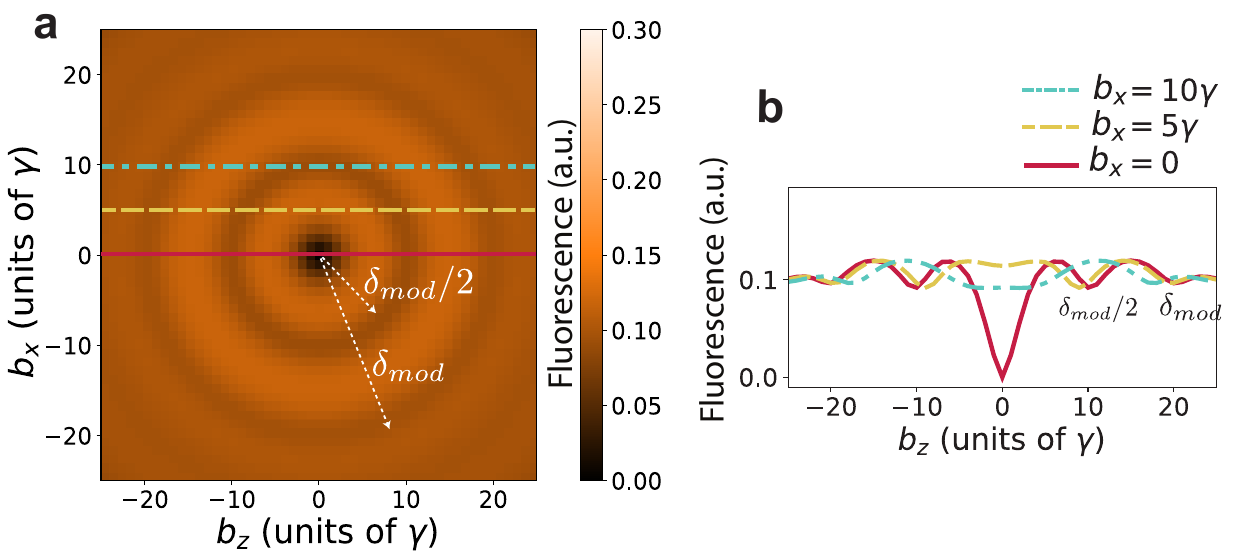}
\caption{\label{fig:contour_bxbz}(a) Computed fluorescence $I_y$ as a function of ($b_x$, $b_z$) showing dark resonances as concentric ``rings" with radii $n\delta_{mod}/2$ ($\theta=\pi/2$, carrier and higher-harmonic sidebands included). Note this is not the same as the spatial image. (b) Horizontal cross-sections at constant $b_x=0, 5\gamma, 10\gamma$ showing reduced contrast at $b_z=0$ due to precession of the atomic dipoles.
} 
\end{figure}

As discussed in the letter, due to the spatial Hanle effect, $B_x$ field causes dipole precession away from the polarization axis ($y$, vertical). This net "tilt" from non-zero $B_x$ allows emitted fluorescence to reach the camera looking down vertically, lifting the fluorescence minimum from zero.
To quantify and visualize this effect, we plotted the vertical fluorescence $I_y$ as a function of $(b_x, b_z) = \frac{g\mu}{\hbar}(B_x, B_z)$ in Fig.~\ref{fig:contour_bxbz}. Here we assume $b_y=0$ and carefully note that the contour plot shows the variation of the fluorescence in the B-field space. It is not the same a spatial image. The contour plot contains a series of dark concentric rings with radii of $n\delta_{mod}/2$ (multiples of $\delta_{mod}/2$), in agreement with the dark-resonance conditions defined in Eq.~\eqref{eq:dark-line-condition}. The horizontal cross-sections at $b_x = 0, 5\gamma, 10\gamma$ show that the minimum fluorescence is no longer zero when $b_x > 0$. 

%
%

%
\begin{figure}[ht]
\centering
\includegraphics[width=0.49\textwidth]{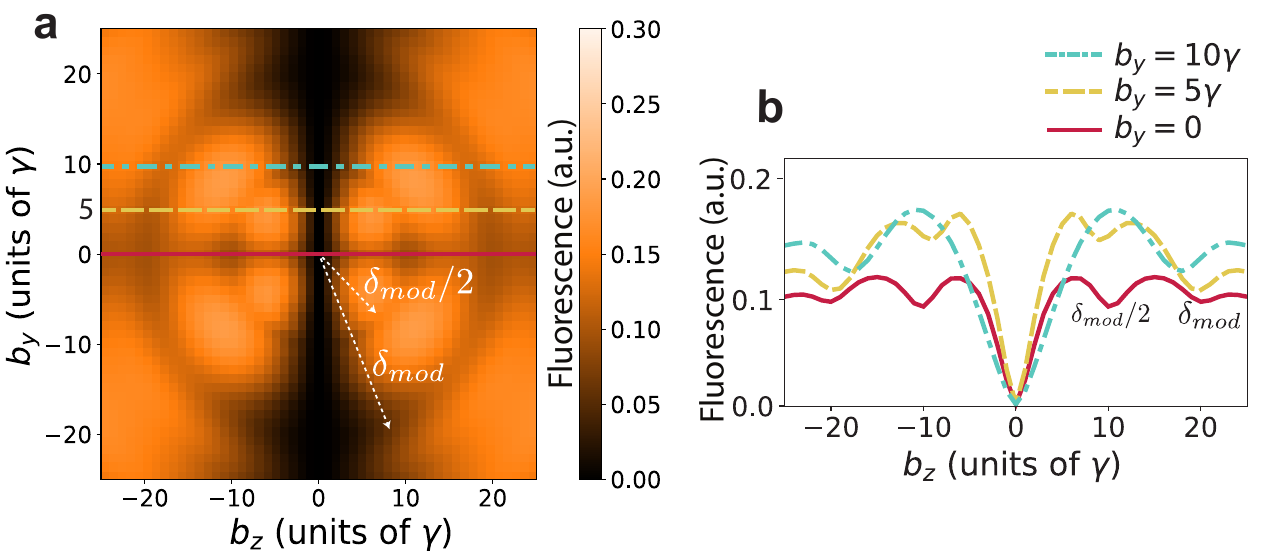}
\caption{\label{fig:contour_bybz}(a) Computed fluorescence $I_y$ as a function of ($b_y$, $b_z$) showing dark resonances as concentric ``rings" with radii $n\delta_{mod}/2$ ($\theta=\pi/2$, carrier and higher-harmonic sidebands included). Fluorescence is zero at $b_z=0$ regardless of $b_y$ values due to the Hanle effect, but there is apparent broadening of the central minimum. Note this is not the same as the spatial image. (b) Horizontal cross-sections at constant $b_y=0, 5\gamma, 10\gamma$ showing apparent shifts in the fluorescence dips.
} 
\end{figure}

On the other hand, we made respective plots for $(b_y, b_z)$ variation in Fig.~\ref{fig:contour_bybz} (assuming $b_x=0$ in this case). The dark-resonance rings are similarly visible, but the main difference here is that the fluorescence $I_y$ remains zero at $b_z=0$ regardless of $b_y$ values, as confirmed by both the contour plot and the cross-sections. This is as expected from the spatial Hanle effect because the $B_y$ field is parallel to the dipole orientation when driven by the $y$-polarized light. The main effects of introducing non-zero $b_y$, as illustrated by the cross-section plots, are broadening of the central minimum and apparent shifts in the other dark-line locations due to the curvature of the dark-resonance rings. 

Both Figs.~\ref{fig:contour_bxbz} and~\ref{fig:contour_bybz} show that in our system, the effects of $B_x$ and $B_y$ are discernible at least when the Yb atoms are driven by $y$-polarized light. 
These distinguishable effects on fluorescence allowed us to estimate $|B_x|$ and $|B_y|$ magnitudes from the experimental data, as demonstrated in Fig.~\ref{fig:B-field-mapping}.
We first infer the plausible range of $|B_x|$ from the reduced contrast (Fig.~\ref{fig:contour_bxbz}) and that of $|B_y|$ from the broadening of the $b_z=0$ minimum (Fig.~\ref{fig:contour_bxbz}). 
For all pairs of $(|B_x|, |B_y|)$ in the inferred range, we computed theoretical fluorescence curves and determined the $(|B_x|, |B_y|)$ values that produced the best agreement with experimental data. 
This is an iterative grid search, in which we compare the number and locations of the fluorescence minima, narrowed down the $(|B_x|, |B_y|)$ range, and recomputed the theoretical fluorescence on a finer grid until we achieved a reasonable fit to the experimental fluorescence data.
This process enables the vector capabilities of our magnetometer. 
However, we cannot discern the signs because of the symmetry in the dark-resonance contours.
\end{document}